# Designing a Communication Bridge between Communities: Participatory Design for a Question-Answering AI Agent

Jeonghyun Lee*

Georgia Institute of Technology, jonnalee@gatech.edu

Vrinda Nandan

Georgia Institute of Technology, vrinda@gatech.edu

Harshvardhan Sikka

Georgia Institute of Technology, harshsikka@gatech.edu

Spencer Rugaber

Georgia Institute of Technology, spencer@cc.gatech.edu

Ashok Goel

Georgia Institute of Technology, ashok@cc.gatech.edu

How do we design an AI system that is intended to act as a communication bridge between two user communities with different mental models and vocabularies? Skillsync is an interactive environment that engages employers (companies) and training providers (colleges) in a sustained dialogue to help them achieve the goal of building a training proposal that successfully meets the needs of the employers and employees. We used a variation of participatory design to elicit requirements for developing AskJill, a question-answering agent that explains how Skillsync works and thus acts as a communication bridge between company and college users. Our study finds that participatory design was useful in guiding the requirements gathering and eliciting user questions for the development of AskJill. Our results also suggest that the two Skillsync user communities perceived glossary assistance as a key feature that AskJill needs to offer, and they would benefit from such a shared vocabulary.

---

*Contact Author

## 1 INTRODUCTION

Human-centered AI is in vogue. Design of human-centered artificial intelligence (AI) typically addresses interaction between a human and an AI agent, for example, a learner engaged with an intelligent tutoring system or a user talking with a conversational AI agent. More recently, human-centered AI has also addressed interaction between an AI agent and a community of users, for example, a virtual teaching assistant for an online class that addresses questions on the class discussion forum [Goel & Polepeddi 2018] or a virtual teaching assistant that introduces students in an online class to one another based on shared identity [Wang, Jing, & Goel, 2022]. Design of socio-technical human-facing and community-facing AI systems has frequently used participatory design as a methodology for engaging various stakeholders and identifying potential problems early in the design process.

However, there is relatively little research on the design of AI agents that need to interact with users from multiple communities. Let us consider, as an illustrative example, an AI agent for linking users from companies seeking employees with specific skills and users from colleges who train potential employees with various skills. Users from the two communities likely have not only differing sets of requirements, but also different mental models and domain vocabularies. How might participatory design help in the design of AI agents that act as bridges between multiple user communities?

The goal of this paper is to illustrate an example of developing an AI system for multiple communities in the domain of reskilling and upskilling of the workforce. We adopted a participatory design and research approach as a methodology to develop AskJill, an AI-enhanced assistive question-answering agent for a newly developed web aplication referred to as Skillsync [Robson et al 2021]. AskJill is designed to offer users real-time assistance in the areas of glossary and task guidance related to the Skillsync interface [Goel et al. 2022]. AskJill can engage an employer or college administrator in sustained dialogues that help bridge communication gaps between the user groups, which will contribute to effective and efficient decision making. Our study aimed to define and refine the requirements of AskJill in SkillSync from a human-centered perspective.

This paper focuses on discussing how participatory design methods guided initial requirements gathering for the development of the AskJill question-answering AI agent specifically designed to bridge the communication gap between employers and educators whose mental models of workforce development differ. Based on the participatory research design framework, we conducted a series of focus groups to gather feedback from a sample of company users and college users. In this paper, we highlighted how focus groups were designed to elicit questions from the two user groups and draw implications with respect to developing a shared vocabulary model between these different communities. Also, the analysis of the focus group data focused on not only categorizing question types but also capturing the voices of users as well as their nuanced interpretations toward the role of AskJill. As a result, we created key requirements and issues for AskJill users, which were taken into consideration throughout the tool development process and design prioritization.

## 2 RELATED WORK

### 2.1 Skillsync for Workforce Development

In the face of fast-paced technological development, demands for upskilling and reskilling hundreds of millions of current US employees have rapidly increased. In spite of increasing demand to promote effective workforce development in the US, there exist issues of a lack of transparency and alignment that threaten collaboration between companies and training providers (i.e., colleges) at scale. Moreover, tackling this issue of workforce training can be challenging because it often requires close coordination among these stakeholders. This unique challenge calls for a need to leverage cutting edge technology such as artificial intelligent (AI) techniques that can enable employers to upskill and reskill their employees effectively and efficiently. In this regard, the Skillsync application [Robson et al., 2022] is designed to help companies address the need to reskill or upskill their employees in partnership with colleges. It also helps colleges match their continuing education and professional development programs to the needs of industry. The application enables companies to document the needed skills in the form of training requests and send these training requests to relevant education providers. It also allows colleges to formulate training proposals in response to the requests based on their educational programs.

Skillsync uses various AI techniques, including machine learning, language models, and matching algorithms, to extract knowledge, skills, and abilities (KSA) [USCDC, 2022; USVA, 2022] from job data and to match them with corresponding courses. The job data originates from sources like the U.S. Department of Labor, industry associations, company job descriptions, and job postings provided by the National Labor Exchange (https://usnlx.com/). The course data is sourced from course catalogs of continuing education and professional development programs at universities and colleges, including technical and community colleges. The extracted KSAs are organized and prioritized in a skills framework. Skillsync helps make the process of matching jobs with educational programs both more efficient and effective. However, the potential adoption and use of Skillsync in companies and colleges requires that it can explain its design and processing, that its processing is transparent, and that its results are trustworthy. These three requirements were taken into account when designing our proposed focus group. We will explain this in more details in the following sections.

## 2.2 Question Answering AI Tools

Question and Answering covers a large variety of different approaches, and there have been several notable works in Q&A using knowledge representations. Predefined rule-based approaches like Bast et al. [Bast and Haussmann, 2015] use templates to extract logic. More recently, Information Retrieval (IR) and Neural Semantic Parser (NSP) based approaches have been proposed [Fu et al., 2020]. The former focuses on extracting entities from natural language questions and makes connections to the knowledge base being covered. Representation learning has been used to great effect in IR based approaches. These methods map question answer pairings into vector space and formulate a matching problem within the distribution of questions and answers. External knowledge has been introduced in the form of structured knowledge bases, web corpuses and the like. Xu et al. [Xu et al., 2016] used Wikipedia as a knowledge base for a KBQA method. Watson [Ferrucci et al., 2010], and thereby Jill Watson [Goel & Polepeddi 201], uses a similar IR representation learning based approach to encode questions into the numerical space of a classifier. AskJill [Goel et al. 2022] evolves from the Jill Watson project.

## 2.3 Participatory Design

Prior research has suggested that participatory design needs to emphasize "hybrid" practices that take place in an "in-between" space that shares spaces from both the users and technology developers [Muller and Druin, 2012]. In such practices, it becomes important that participants experience negotiation, co-construction of understandings, and collective discovery to create new ideas. Regarding how to facilitate the participatory design process, multiple facets of the participatory design facilitator role have been identified, including (1) trust builder, (2) enabler, (3) inquirer, (4) direction setter, (5) value provider, and (6) users' advocate [Dahl and Sharma, 2022]. In particular, the inquirer role is responsible for challenging participants' assumptions through asking follow-up questions and inviting participants to express and share conflicting values and perspectives with one another. Additionally, recent discussions about the application of participatory to AI systems address a potential barrier for facilitating collaboration among co-designers, users, and stakeholders due to differing levels of familiarity with and/or background knowledge about AI [Hossain and Ahmed, 2021; Zytko et al., 2022].

Taking these perspectives into consideration, our participatory design research explored the process of how different end-user groups came to work together and collectively discover insights to build a shared model in an AI system. Specifically, this study was guided by Spinuzzi's [Spinuzzi, 2005] three stages of participatory design. These stages include initial exploration (Stage 1), discovery processes (Stage 2), and then prototyping (Stage 3). Our work was primarily guided by the initial exploration of work stage that involves various qualitative methods such as observations, interviews, and walkthroughs. The study aimed to draw human-centered implications mainly from findings of focus group activities that took place during the discovery processes stage.

## 3 PARTICIPATORY DESIGN OF ASKJILL

### 3.1 Issues and Hypotheses

Our work was guided by a set of hypotheses to address two general issues in designing a data-driven and human-centered question-answering system targeted toward multiple communities. The first issue is figuring out how to build such a question-answering system that focuses on providing support for vocabulary needs from different communities. To address this issue, we posited that adopting participatory design would be a viable approach to eliciting questions from multiple user groups. The second issue is how to implement participatory design to the development process that involves multiple user groups. Our corresponding hypothesis is that inviting the user groups to focus group sessions that included intentional breakout group discussions and a whole group debrief activity would be an appropriate software implementation strategy. The following sections will provide further details of how our study illustrates this participatory design process.

### 3.2 Participatory Design Process

The ultimate goal of our research study is to define the requirements of AskJill in SkillSync [Polepeddi and Goel, 2017; Goel et al., 2022] from a human-centered perspective. The research team has so far performed several iterations of participatory design and requirements gathering of the AskJill AI agent since the launch of the project in Fall 2020. As illustrated in Figure 1, we conducted initial focus groups to elicit user questions and feedback, which became the basis of gathering functional and procedural requirements in the following year. This iterative process allowed the researchers to develop the target software product based on the user-centered participatory design [Spinuzzi, 2005]. The requirements gathering process through the focus group study enabled us to develop such key features as glossary assistance.

AskJill in SkillSync was developed based on a Task-Method-Knowledge (TMK) model [Murdock and Goel, 2008; Goel and Rugaber, 2017] of SkillSync to build machine readable knowledge representations that are also interpretable by human users who interact with the AskJill. A TMK model is conceptualized as being compositional, causal, and hierarchical, which enables an AI agent to generate causal explanations of how it works by encoding information in three different ways. Specifically, tasks represent the "Why?" of a system, specifying the goals of the AI agent. Methods represent the "How?" of a system, describing the internal processing of the agent. Knowledge captures the "What?" of the AI agent, expressing the information that the agent is operating on. Next, we conducted a second focus group study in Spring 2022 and findings of the focus group informed the development of features based on task and knowledge models and plus the design of a method model that is considered as a more complex feature compared to the previously developed features.

Across the participatory design process, target users mainly consisted of company-side and college-side partners who were interested in interacting with and potentially utilizing the SkillSync platform. The SkillSync user interface for each user-side differed by a specific goal and its related tasks (see Figure 2 for key component tasks). In general, company users were asked to provide detailed information about training to upskill or reskill their employee trainees. The training request was then sent to college users for review and they were asked to create a training proposal to meet the company needs.

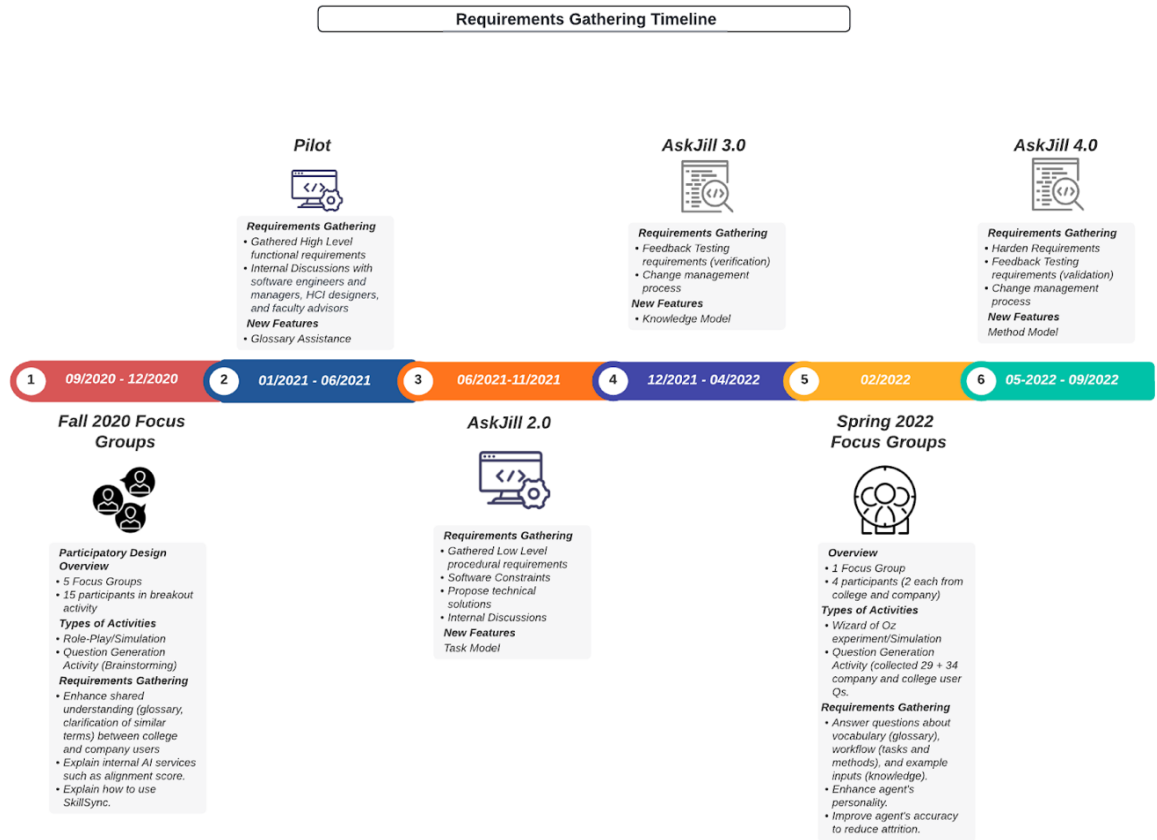

Figure 1: AskJill participatory design requirements gathering timeline.

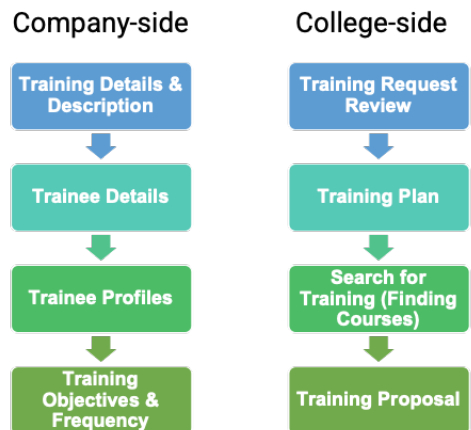

Figure 2: Key component tasks of each user-side in the SkillSync interface

### 3.3 Scope of This Paper

Although both of the SkillSync user-sides (i.e., company and college) share a common goal of providing employees with the most effective training opportunities, it is important to note that they might bring different understandings or interpretations towards the same terminology due to potential differences in their situated professional and cultural context, including language or jargon (e.g., whether and how to distinguish terms like 'skills' and 'credentials'). In fact, while existing AI systems are typically designed to target a single community of users, AskJill intends to bridge between these two communities [Goel et al. 2022]. From the software perspective, building shared vocabulary is considered as a steppingstone to developing other complex features [Goel and Rugaber 2017]. Also, the TMK model framework suggests that users' vocabulary manifests their knowledge representations about fundamental concepts that are being modeled [Murdock and Goel, 2008; Goel et al., 2022].

Our study was broadly guided by two research questions. How can participatory design best be used when multiple classes of users are involved? Which specific method can be used to execute participatory design with multiple user groups? In this paper, the authors will focus on deeply examining the initial stage of the participatory design process during which the researchers conducted a series of focus groups. Considering that the primary goal of SkillSync is to enhance communications between the two user groups, this paper aims to illustrate how participatory design allowed us to elicit user questions and understand needs around shared vocabulary between company and college users.

As outcomes of participatory design, we hypothesized that provision of a glossary assistance feature will emerge as one of the target features that will potentially benefit AskJill users from both company and college groups. Specifically, we conducted a focus group study as our primary methodology for participatory design research. The focus groups were designed to examine the extent to which the Skillsync user groups perceive glossary assistance as a key feature that AskJill needs to offer. Finally, based on the insights that we gathered, we extracted a set of fundamental requirements to develop AskJill that is capable of generating a shared vocabulary model for the two SkillSync user groups.

## 4 INITIAL FOCUS GROUPS

### 4.1 Methods

*4.1.1 Participants and Settings.*

A total of fifteen people including four men and eleven women volunteered to participate in one of five focus groups that took place via Zoom video conferencing (the focus groups met virtually because of the Covid-19 Pandemic). The focus groups spanned approximately three months during Fall 2020. The participants consisted of ten employers who came from three companies across the country (i.e., company users) and five administrative staff or educators from three different higher education institutions (i.e., college users). The two participant groups were introduced to different tasks and interfaces in the Skillsync platform. That is, company users were mainly asked to provide detailed descriptions of target trainees (e.g., current skills or competencies) and training that they would request from a college provider. On the other hand, college users were asked to review a training request submitted from a company user and then build a training proposal that includes a list of courses that the college can offer to best meet the training needs.

*4.1.2 Procedures.*

Each focus group began with a 10-minute introduction to the concept and prototype of SkillSync, followed by another 10-minute high-level discussion about the process of creating training requests and training proposals for upskilling or reskilling current employees. Next, the participants were invited to a breakout group activity that placed company participants in one group and college participants in another group. The breakout group activity lasted for about 45-50 minutes. Each respective group interacted with an interface customized for the needs of that group. Using the SkillSync prototype interface, the focus group facilitator showed a demo of (1) how a company requests a training project and (2) based on this information how a college builds and shares a proposal with the company. Participants watched the facilitator navigating the prototype through the shared screen. The facilitator paused at several landing pages to elicit participants' feedback about what questions they would ask of AskJill regarding how to use the SkillSync interface (e.g., "What questions come to your mind when you look at this page?"). Based on a think-aloud technique [Spinuzzi, 2005], participants were asked to say out loud any questions and thoughts that arose as they explored the prototype. The primary goal of this question-generation activity was to uncover salient types of questions that users would ask of an AI agent. As a result, we were able to extract 62 questions that were explicitly raised by the participants, which became the basis of our analysis.

After the breakout group session, the company and college participants were merged back into the main meeting room for an integrative debrief session where any remaining questions or comments could be shared and recognized. During the debrief activity, the participants were asked to be cognizant of vocabulary and raise a question if someone used a term that they were not sure about. Including this activity to each focus group session was important to gain insights to essential elements for building a shared model for vocabulary or knowledge representations [Goel et al. 2022] between the two user groups.

*4.1.3 Data Analysis.*

The purpose of the focus group activities was to draw inferences about the possible usage of AskJill. We focused on capturing notable questions or issues that participants addressed while they were interacting with the prototype. Additionally, we conducted descriptive analysis of the questions that emerged from the participants. To analyze collected qualitative data, we focused on generating broad themes and categories of user questions and identifying a set of emerging requirements to design AskJill throughout the iterative analysis process.

Specifically, in order to inform development and training of the knowledge base for AskJill, user questions that we extracted from the focus group data were categorized based on the TMK model framework. As a conceptual guide for this framework, an explanatory ladder [Goel et al. 2022] was introduced to the process of analyzing the elicited user questions. The explanatory ladder consists of three main categories of explanations, where the depth of explanations increases as we go up the ladder from the category of vocabulary to knowledge and to reasoning. That is, at the lowest step of the ladder, the agent can answer questions about vocabulary; at the next step, questions about knowledge in two subcategories (raw data and information, and inferred knowledge); and at the highest step, questions about reasoning with three subcategories (context of the current interaction, task, the process to accomplish those tasks).

*4.1.4 Results.*

**Categorization of User Questions and User Insights about Vocabulary Models.** As a result of the focus groups, a total of 62 unique questions were explicitly raised by the participants, with 36 questions being relevant to the company-side interface and the remaining 26 relevant to the college-side interface. The questions were collected and bucketed based on the SkillSync dashboard that was displayed during the demos for the breakout group activity. These questions were carefully reviewed and categorized based on three categories and their corresponding subcategories in the TMK model's explanatory ladder framework. The main categories included vocabulary, knowledge (including inferred knowledge, raw data and information), and reasoning (including context, process, and task). Overall, the percentages of each category and subcategory appear to be similar between the company and college participant groups (Figure 3). Yet, it is notable that there was a higher percentage of vocabulary-type questions within the college group (23%), compared to the company group (19%). While both participant groups commonly asked task-type questions, the percentage was higher in the company group (28%) than the college group (23%). When the elicited questions from each group were combined, 20% of the total questions across both user-sides were vocabulary-type questions, 25% were knowledge-type questions, and the remaining 56% were reasoning-type questions (see Table 1 for example questions).

Our classification results revealed that both sides of users actively generated task-related questions to seek clarification or guidance for successful task completion in the SkillSync interface. These questions generally pertained to user experience and application specific functionality. Also, the participants commonly asked questions about sources of information and inferred meaning of tasks or concepts provided in the SkillSync platform. However, despite the vocabulary questions having the lowest percentage of the total user-generated questions, it is noteworthy that many of these questions provided fundamental information that is required to build a shared vocabulary model for the two end-users. In the next section, we will provide further details and examples that illustrate how the focus group allowed the company and college participants to discover common needs for AskJill and discuss issues that need to be resolved in order to establish a shared understanding toward vocabulary in SkillSync.

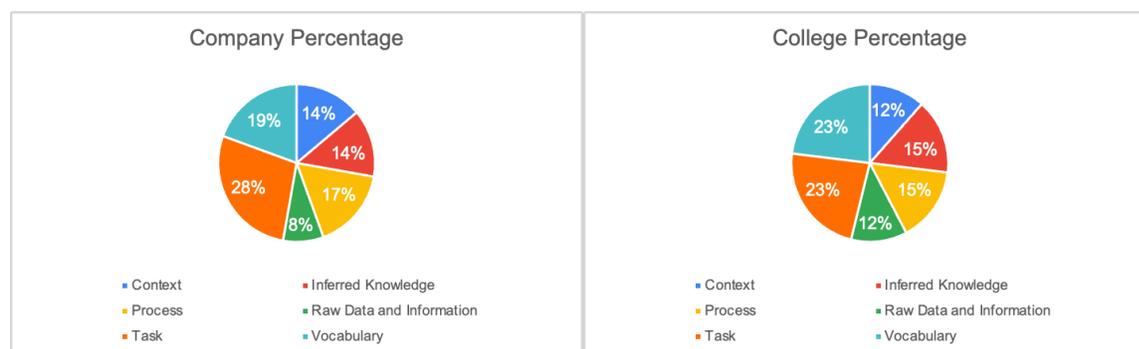

Figure 3: Comparison of User Question Categorization Results between Company and College Group

Table 1: Categorization and Illustrative Examples of User Questions

| Question Category & Sub-Category | Percentage of Total Elicited Questions | Example Questions |
|---|---|---|
| **Vocabulary** | 20% | <ul><li>How do you define "trainees"?</li><li>How are KSAs/competencies/tasks different?</li><li>How do "skills" and "credentials" differ?</li></ul> |
| **Knowledge** | 25% | |
|     Inferred Knowledge | 15% | <ul><li>What does alignment score tell us about?</li><li>How do I interpret symbols or icons on the screen?</li></ul> |
|     Raw Data and Information | 10% | <ul><li>Where is Target Occupational Tasks coming from?</li><li>Whose courses am I searching now?</li></ul> |
| **Reasoning** | 56% | |
|     Context | 13% | <ul><li>What is the timeframe that company would need?</li><li>Who are my potential providers?</li></ul> |
| Process | 16% | • What should I do if what I am looking for is not listed in the search results?<br>• When can I expect to receive the proposal back? |
| Task | 26% | • How do I filter training course results?<br>• How do I add more details about course logistics?<br>• How do I select and add occupational tasks? |
| Total | 100% | |

*Outcomes from the Focus Group activities.* Within the elicited vocabulary-type questions, both company and college participants often asked questions around domain knowledge to clarify the definition of similar terms that appeared in the SkillSync platform. For example, a common question was how KSAs (knowledge, skills, abilities) differed from related concepts such as tasks, competencies, and credentials. In fact, these terms can be used interchangeably, and it is even possible that they are defined differently depending on the context. For example, one company participant pointed out that 'competencies' would be interpreted from the perspective of business management. On the other hand, college users might view it as a pedagogical concept. Some participants even raised an issue regarding the selection of certain keywords, as illustrated in the following quote: "It is interesting that it is 'tasks' rather than skills or competencies [under Training Objectives]."

Other frequently asked questions were about seeking definitions of ambiguous terms used in the titles or task prompts that appeared in each landing page, as one company participant mentioned, "I want some definition underneath these [terminology for training category]. Such examples included 'trainee', 'ongoing training', 'objectives', and 'alignment score'. These examples illustrate that there still remains an unmet need to enhance shared understandings, that is, whether terms are transferable or "standardized" between company- and college-side users.

We observed that the whole-group debrief activity, which took place immediately after the breakout group discussion, helped the company and college participants to compare how they would use and interpret not only those terms that appeared in the SkillSync platform (e.g., KSA) but also other terms associated with the workforce training situation (e.g., credentials). Furthermore, this debrief activity appeared to enable the two participant groups to uncover converging needs for AskJill centered around glossary assistance, as shown in Table 2. Indeed, an ultimate goal for both groups would be to build a training proposal that successfully meets the needs of future trainees. To achieve this goal, clear and accurate mapping of proposed training courses and skills that will be taught in those courses is essential to achieve the goal. Establishing alignment between language from company (client) and that from college (provider) is crucial during this process. In that regard, the design of our focus group study can be useful for future participatory design of a shared vocabulary model in a question-answering AI agent for multiple end-user communities.

Table 2: Converging Need for AskJill Based on Elicited Vocabulary-Type Questions

| Example Vocabulary-Type Questions Elicited from Company Users | Example Vocabulary-Type Questions Elicited from College Users | Converging Need for AskJill |
|---|---|---|
| • What is the meaning of a selected KSA?<br>• How do "skills" and "credentials" differ?<br>• How are "KSAs" different from tasks?<br>• How are KSAs/competencies/tasks different?<br>• What does it mean to be one time vs ongoing training?<br>• How do you define "trainees"? | • How do you define KSA, occupational tasks, skills, competencies?<br>• What is the difference between "existing skills" and "targeted skills"?<br>• Can I find global descriptions on the title of each landing page (e.g., Trainee Profile, Alignment Details, etc.)?<br>• Are objectives different from outcomes? | • Need for clarifying the definition of terms that have the similar meaning as KSA to enable users to understand training needs and build proper training content.<br>• Need for providing glossary for vocabulary on the SkillSync interface to enable users to articulate training details and logistics. |

***Other Insights from User Questions.*** In addition to the analysis of the user questions that are directed to AskJill, we examined other related comments made by the company and college participants to deepen an understanding of the unique needs of each user side (i.e., company and college). This provided the research team with important information that needed to be considered especially when developing non-functional features of AskJill such as efficiency and maintainability.

First, the participants commonly mentioned the need for assistance with efficient information processing by which their cognitive load can decrease. Because the SkillSync interface often displays, as part of the task process, text-heavy information that came from extensive course or job catalogs, participants occasionally reported that they felt overwhelmed about the quantity of the given information. For instance, users are asked to scroll down a long list of potential courses along with their descriptions in order to choose the best courses that meet their training needs. Therefore, participants seemed to feel that it would be challenging to quickly determine which information is more important than others for effective decision making. Several participants explicitly asked how AskJill will be capable of handling this problem, as seen in the following sample quote, "Is there a way to make a targeted decision without reading through all of the search results? How will this [AskJill] be able to assist with finding targeted information?"

Next, we often noted that the participants expressed task-oriented needs. For instance, company participants sought a guide to articulate their training needs in their training requests while college participants wanted to gain more confidence in estimating a degree of match between company needs and what they are able to offer. For many company participants, their inquiries focused on wanting to know how to best describe desired skills or tasks that their employees need to be trained on. Regarding college participants, they were specifically interested in the notion of "alignment score", which is an underlying tool of the SkillSync interface that indicates the degree of matching between specific training needs that employers specified in a training request and course content that is proposed in a training proposal. This finding implies a possibility that participants view the role of AskJill as beyond that of an assistance tool and potentially as "a coach" that can provide task-specific feedback on whether and how the user can successfully build a training request or proposal.

***Gathering of Design Requirements of AskJill in SkillSync.*** Informed by the focus group findings, we followed a user-centered participatory approach during the requirements gathering process by asking ourselves "What should AskJill be able to do to enable the best user experience?" This question triggered conversations both with the end users (during the focus groups through the participatory design process) and within our internal cross functional team of software engineers and managers, HCI designers, and researchers (face-to-face conversations, email discussions, past knowledge). As a result, we were able to synthesize outcomes from the participatory design research and past results from researching question-answering agents. Then, we analyzed the resulting data and built consensus around the high-level functional requirements and performance requirements that we should prioritize at each stage of the development process and their desired user impact. Armed with this information, we documented the first requirement set.

Next, we extracted the fine-grained procedural requirements while evaluating the inherent software constraints and their implications. This led to further requirement discussions and helped us propose technical solutions that fulfilled the revised set of requirements. Finally, we communicated the resulting requirements and their impact to the entire team and built a shared understanding around the envisioned end product.

The requirements gathering process continued during the software development process, as we continually evaluated existing requirements and fed back new information from verification and validation through end user focus groups consisting of live training demos and user interaction with SkillSync. Also, we iteratively revised our requirements by either simplifying or expanding them through a change management process. The requirements also served as a sounding board (i.e., Is the requirement satisfied or not satisfied?) for us to judge the completeness of the software development

Designing a Communication Bridge between Communitiesprocess. Table 3 shows the Final Requirements for Pilot and Table 4 presents example template questions that were created based on the extracted requirements.

Table 3: Final Requirements for Pilot

| Index | Requirement Category | Requirement Sub-Categories (Implemented in Actual Pilot) |
|---|---|---|
| 1 | AskJill provides a user with glossary assistance upon request in a prompt manner. | 1A. AskJill will be able to answer specific questions about glossary from a user, either a company- or college-side, of the Skillsync platform.<br>1B. AskJill should provide a user with the definition of a requested term in a complete sentence(s).<br>1C. AskJill allows a user to ask about one term at a time.<br>1D. AskJill should provide a response to a user within .5 milliseconds.<br>1F. AskJill is designed to handle error(s) in a user question (e.g., typos, out-of-scope questions). |
| 2 | AskJill enhances clarity of important vocabulary or features embedded in Skillsync. | 2A. AskJill provides users with a nuanced definition of a similar set of terminology related with KSAs and occupational tasks (e.g., skills, competencies, objectives, outcomes). |
| 3 | AskJill is designed to reduce the cognitive load of SkillSync users and therefore help them process text-heavy information in an effective manner. | 3A. The maximum length of AskJill's single text response should be 160 characters.<br>3C. AskJill will serve as a virtual assistant that can be easily recognized and available throughout the entire process during which users interact with SkillSync. |

Table 4: Example Template Questions

| Index | Template Questions | Index | Template Questions |
|---|---|---|---|
| 1 | What is _____ | 13 | I am not sure what _______ is |
| 2 | What are _____ | 14 | Can/Could you clarify what _____ is |
| 3 | Who are _______ | 15 | Can/Could you explain what ______ is |
| 4 | What does ______ mean | | Can/Could you give an explanation for ______ |
| 5 | What does ______ stand for? | 16 | Can/Could you define _______ |
| 6 | What is the definition of ______ | 17 | Can/Could you describe _______ |
| | What is _____ definition? | 18 | Clarify ______ |
| 7 | How does Skillsync use _______ | 19 | Explain ______ |
| | How does ______ work | 20 | Define ______ |
| 8 | What is a _____ in Skillsync | 21 | Describe ______ |
| 9 | Why is it asking for _______ | 22 | I have no idea what _____ means |
| 10 | I don't know what _____ is | 23 | What do you mean by _____ |
| 11 | I don't know what _____ means | 24 | What if I'm unsure what ______ is |
| 12 | I am confused by _______ | 25 | What should I do with ______ |

## 5 DISCUSSION AND CONCLUSION

Guided by participatory design, our study explored a method to elicit human-centered questions in order to define the requirements of the AskJill tool. Participatory design enabled us to capture the voices of AskJill users from two communities, company and college, mainly through observations and analysis of their verbal and behavioral feedback while each side of the users was interacting with different SkillSync interfaces (either directly or indirectly). By using the think-aloud protocol technique [Spinuzzi, 2005], we had an opportunity to explicitly bring their awareness and attention to AskJill prior to the pilot experiment. This allowed researchers to gain knowledge on what questions come to an AskJill user's mind and issues that need to be addressed, which subsequently improved an understanding of specific needs and challenges in human-AI interactions. Despite logistics barriers and meeting requirements, participatory design effectively drew requirements about the ultimate design and implementation of the system.

Our findings provide a valuable insight into how participatory design can be useful to facilitating a process of generating a shared vocabulary model for multiple user groups, as one of the essential components of a question-answering AI agent tool. As illustrated in Figure 2, the context of our study is unique in that the participants came from different professional backgrounds with different mental models and vocabularies related to reskilling and upskilling. Considering that a vocabulary is the basis for representing and reasoning about the world, it would normally be expected that an AI system like AskJill runs two distinct vocabulary models for each company- and college-side. *However, our findings suggested that AskJill in the context of SkillSync could potentially function as a shared mental model when it comes to providing glossary assistance to both user sides. Such a shared vocabulary model would facilitate transparent and effective communications between the two user groups.*

This study also illustrated one viable approach to implement participatory design to gather initial user requirements for an AI system that is capable of bridging two user communities with different vocabulary models by providing a glossary of terms related to a software platform. In our case, focus groups were designed to engage the company and college users with an active question-generating activity with other peer users, which was supplemented by a debrief activity in which both user groups gathered and had an opportunity to clarify vocabulary and discuss any remaining issues that need to be resolved to improve their interaction with the SkillSync platform. As one important takeaway from our study design, the debrief sessions can be viewed as a venue to explore and identify converging needs between multiple SkillSync user communities around glossary assistance, which was crucial in the prioritization of requirements when designing AskJill.

Furthermore, our findings showed that participatory design allowed the participants to observe the SkillSync interface and raise explicit questions that reflected the constraints and affordances of AskJill within the interface. Moreover, as our participants represented each of the two user sides, the types of questions became richer and contextualized. It is noteworthy that the debrief session during the conclusion of each focus group session led us to obtain a "process" outcome. In other words, the participants had an opportunity to communicate their needs to other users with different professional backgrounds and compare each other's perspectives during the focus groups.

There are several limitations that require cautious interpretation of our study findings. First, some of users' immediate needs may not necessarily be prioritized in the tool development due to time and resource constraints. For example, some users struggled with how to complete proposal building/request tasks rather than clarifying definitions of terms; while we initially focused on the glossary assistance function of AskJill for the pilot. Additionally, this study captured only a partial snapshot of our research team's participatory design process. It was out of scope of our paper to discuss how we iterated the design process and how user interactions and feedback evolved. However, as we progressed beyond the initial requirement gathering task, we observed that users continuously provided nuanced insight into software constraints and areas of improvement in the design of AskJill. Also, as users became able to ask questions in the AskJill application, these user questions were logged and converted into user stories, resulting in new features and enhancements. Our future study will highlight how the presence of an intelligent agent like AskJill enables a design and development team to continuously engage in the participatory design process, even after deployment.

**REFERENCES**

[Bast and Haussmann, 2015] Hannah Bast and Elmar Haussmann. More accurate question answering on freebase. In Proceedings of the 24th ACM International on Conference on Information and Knowledge Management, pages 1431–1440, 2015.

Designing a Communication Bridge between Communities